\documentclass[12pt]{article}
\usepackage{epsfig}

\begin{document}

\begin{center}

\bigskip
V.A.Sadovnikova

\bf{On the relation between the long-wave instability and the pion
condensation in nuclear matter}

\bigskip
Abstract
\end{center}

The solutions to the zero-sound frequency equation and pion
dispersion equation are considered for the different values of the
parameters of the effective particle-hole interaction.
The branches of solutions, responsible for the long-wave
instability (Pomeranchuk,\cite{1}) of the nuclear matter $\omega_P(k)$
are presented for the both equations. It is shown that in the pion
dispersion equation the solutions $\omega_P(k)$ result in the
instability not only for values of the spin-isospin constant $g'\leq
-1/2$, violating stability criteria, but for all $g' < g'_p$
($|g'_p|\ll 1$ and $g'_p < 0$). At $g' > g'_p$ the branches
$\omega_P(k)$ turn into the solutions responsible for pion
condensation.

\begin{center}
{\bf Introduction}
\end{center}

In this paper the various excitations in symmetric nuclear matter are
 investigated. The nuclear matter  is considered as a normal
 Fermi liquid  \cite{LP,PN}.  The quasiparticle-quasihole interaction
is taken in the form \cite{AB1,AB3}

\begin{equation} {\cal F}(\vec k_1,\vec \sigma_1,\vec
\tau_1;\vec k_1,\vec \sigma_1,\vec \tau_1)=C_0( f(\vec k_1,\vec k_2)+
f'(\vec k_1,\vec k_2)(\vec \tau_1\vec \tau_2)+ g(\vec k_1,\vec
k_2)(\vec \sigma_1\vec \sigma_2)+ \end{equation} $$ g'(\vec k_1,\vec
k_2)(\vec \sigma_1\vec \sigma_2)(\vec \tau_1\vec \tau_2)),
$$
where $\vec\sigma, \vec\tau$ are the Pauli matrices in spin and
isospin space, respectively;  $C_0=N_0^{-1}$ while $N_0=p_Fm/\pi^2$
 is the state density of the one sort particles on the Fermi
surface.  The single particle momenta
$\vec k_1$ and $\vec k_2$ are at the Fermi surface. In the homogeneous
medium the functions  $f, f', g, g'$ depend on the angle
$\theta$ between $\vec k_1$ and $\vec k_2$ only.  This permits to
expand them in Legendre polynomials: $f(\vec k_1,\vec k_2)=\sum_{l}
f_lP_l(cos \theta)$ and likewise for $f',g,g'$

Stability conditions of the ground state
with respect to the long-wave excitations were obtained in the paper
\cite{1}.  The spherical Fermi surface  is stable against the
small deformations if the parameters in (1) are restricted by
certain conditions  \cite{1,AB1}. Say, for the parameter $f_l$ it is

\begin{equation}
\label{3} \frac{2f_l}{(2l+1)} > -1.
\end{equation}
There are  analogous expressions for the functions $f', g, g'$.

In this paper the simple model of interaction (1) is used with the
functions $f(\vec k_1,\vec k_2)$ assumed to be the constants
\cite{LP,PN}.  Hence only the terms with $l=0$ have nonzero values
 in the Legendre expansion.  Below we omit index $0$. In this case
the stability criteria are reduced to $f > -1/2$. The same is true for
others functions.

The frequencies of the zero-sound collective excitations are the poles
of the full scattering amplitude $\Gamma$ in the matter \cite{LP,AB1}.

\begin{equation}\label{4} \Gamma={\cal F} +
 {\cal F} A\Gamma, \quad \Gamma=\frac{{\cal F} } {1-{\cal F} A}.
\end{equation}
 Here $A$ is the sum of the two particle-hole loops.
One of them corresponds to the excitation of the particle-hole pair
and the second one to the absorption.
The zero-sound frequency equation
\begin{equation}\label{5} 1 - {\cal F} A(k,\omega,p_F) = 0
\end{equation}
can be written for  any type of excitations: scalar, spin, isospin,
spin-isospin.

In sect.1 we discuss the origin of the ground state instability
from the point of view of  solutions to (\ref{5}).  It is known
\cite{PN,T} that at $f\leq -1/2$  equation (\ref{5}) has
two imaginary solutions. It is shown further how they appear on the
physical sheet of the complex plane of $\omega$, where they are at
$f>-1/2$. The character of instability related to this solutions is
considered. We follow the evolution of the instability while
including  the pion-nucleon interaction.

Of course at $f\leq -1/2$ we obtain the negative compressibility
of matter \cite{LP,PN}. This fact and the
 appearance of the imaginary solutions on the
physical sheet  are the signals of the phase transition.
Thus, there are both the dispersion equation and its imaginary
solutions at  $f\leq -1/2$. However, there is no the physical object
described by them.
Nevertheless, the investigation of these solutions
can give a useful information about the physical processes in matter.

Equation (\ref{5}) contains several parameters which
 characterize the matter. These are the Fermi momentum $p_F$,
the effective nucleon mass $m$ and the constants of the effective
interaction (1).  We fix the value of $p_F$ and $m$: $p_F$=268 MeV
(this corresponds to the equilibrium density $\rho_0 = 2 p_F^3/3 \pi^2
= 0.16$fm$^{-3}$) and $m=0.8 m_0$, $m_0$=0.94 GeV. The aim of this
paper is to investigate the dependence of solutions to the zero-sound
equation on the value of $f$.  In the simple model of the quasiparticle
interaction the results for the all the types of zero-sound excitations
are the same \cite{LP,PN}.

We demonstrate that
there are the various branches of excitations on the physical sheet
for the different intervals of the values of $f$. Namely,
1)~at  $f > 0$ the  zero-sound excitations $\omega_s(k)$ exist,
2)~at $-1/2 <f < 0$   solutions are not found,
3)~at $f \leq -1/2$  a purely imaginary branch $\omega_P(k)$ of
solutions appears. For this branch we find $\omega_P(k)^2 \leq 0$ and
 $\omega_P(k=0)=0.$
At $-1/2 <f < 0$ $\omega_P(k)$ is placed on the
unphysical sheet.
The branch $\omega_P(k)$  is responsible for the
instability of the ground state with respect to the long-wave
excitations.

Let us pass from the zero-sound dispersion equation to pion dispersion
equation and consider the waves with quantum numbers  $J^\pi=0^-$.
Following Migdal \cite{AB1,AB3}, we include the additional interaction
 corresponding to  the one pion exchange into (1)
\begin{equation}\label{5.1}
 g'_t =  g' + \frac{1}{C_0}\left(\frac{f_{\pi
NN}}{m_\pi}\right)^2 \frac{k^2}{\omega^2-m_\pi^2-k^2}.
\end{equation}

Replacing $g'$ by $g'_t$ in the expression (1)  and using Eq.(\ref{5.1})
we obtain after some
transforms a pion dispersion equation
\begin{equation}\label{6}
\omega^2-m_\pi^2-k^2 -\Pi(k,\omega,p_F)=0,
\end{equation}
$$
\Pi(k,\omega,p_F)=-\left(\frac{f_{\pi NN}}{m_\pi}\right)^2
\frac{k^2 A}{1-C_0g'A}.
$$
Here $\Pi(k,\omega,p_F)$ is a pion polarization operator in nuclear
matter renormalized by the spin-isospin particle-hole
interaction $C_0g'$.

To simplify the qualitative picture, the isobar-hole intermediate
states and scalar part of the polarization operator are not taken into
account. But they will be taken into account in several points and this
will be indicated specially .  The poles of the polarization operator
are the solutions to (\ref{5}).  Eq.(\ref{6}) has the solutions which
correspond to $\omega_s(k)$ and $\omega_P(k)$. We denote them by
$\omega^\pi_s(k)$, $\omega^\pi_P(k)$. The correspondence is
 established by the condition that
the branches $\omega^\pi_s(k)$ and $\omega^\pi_P(k)$ approach to the
poles of polarization operator   $\omega_s(k)$ and
$\omega_P(k)$ at $f_{\pi NN}\rightarrow 0$  .

Our calculations demonstrate that the pion-nucleon interaction
distorts the zero-sound branch $\omega_s(k)$ weakly . However the
behaviour of $\omega_P(k)$ is changed significantly and we obtain
instability of the ground state not only at $g'\leq -1/2$  but
also at $g'_p > g' > -1/2$ ($|g'_p|\ll$ 1, $g'_p < 0$). At $g'_p > g' >
-1/2$ the imaginary branch $\omega^\pi_P(k)$ appears on the physical
sheet at the critical density $\rho_c$  at some value of $k_{c}$ for
which $\omega^\pi_P(k_{c})=0.$ Momentum $k_{c}$ is changed from
$k\sim p_F$ at $g'\sim 0$ to $k\rightarrow 0$ at $g'\rightarrow -1/2$.
At $\rho > \rho_c$ the solutions $\omega^\pi_P(k)$ are on the
physical sheet for the certain values momenta. The value of
$\rho_c$ is determined by the parameters of matter  $p_F, m^*, g'$.

In sect.2 the solutions to the pion dispersion equation are analyzed.
They were regarded in detail in relation to the pion condensation.
The results discussed here are complementary to those, obtained in the
well-known papers \cite{AB3,EW}.  The pion
dispersion equation has the following branches of solution:
1)~the pion branch $\omega_\pi(k)$, at $f_{\pi NN}\rightarrow 0$
it is simply $\omega_\pi^2(k)= m_\pi^2 + k^2;$  2)~the isobar branch
$\omega_\Delta(k);$ 3)~the zero-sound $\omega^\pi_s(k).$ Besides there
is also 4)~'condensate' branch $\omega_c(k)$ responsible for the pion
condensation \cite{DRS,SR}. It comes from the same point as
$\omega_\pi(k)$, $\omega_c(k=0)=m_\pi$ and moves over the unphysical
sheet. It  goes over to the physical sheet at the density being larger
 then a critical one $\rho \geq \rho_c$ at certain value of $k=k_1$ and
 returns to the same unphysical sheet at $k=k_2$.  The momenta $k_1$
 and $k_2$  depend on the density: at the critical density
 $k_1=k_2$.  On the physical sheet $\omega_c(k)$ is imaginary
 $\omega_c^2(k)\leq 0$. We see that the behaviour of $\omega_c(k)$ is
the same as $\omega^\pi_P(k)$, moreover these branches are  on
the same unphysical sheet, starting, however, from  different points.
They come to the physical sheet at different values of $g'$.

 It is shown in sect.3 that for the any value of $g'$ (the
values $g' \leq 1$ were considered)  there is such a value of the
matter density $\rho_c$ that for $\rho \leq \rho_c$ the  nuclear matter
is unstable.  When $g'$ is changed near zero instability with respect
to $\omega^\pi_P(k)$ turns into  instability with respect to
$\omega_c(k)$. This replacement is not apparent on the physical sheet.
However  this is seen clearly on the unphysical sheet since these
branches start in different points:  $\omega^\pi_P(k=0)=0$ and
$\omega_c(k=0)=m_\pi$.  The interchange of $\omega^\pi_P(k)$ and
$\omega_c(k)$  is the result of interaction of the branches on the
unphysical sheet:  at some  $g'=g'_{p}$ ($|g'_{p}|\ll 1$, $g'_{p} < 0$)
the branch $\omega_c(k)$ and symmetric $\omega^1_c(k)$ block the branch
$\omega^\pi_P(k)$ on the imaginary axis of the unphysical sheet.  At
$g'<g'_{p}$  the branch $\omega^\pi_P(k)$ goes  to the physical sheet
but at $g'>g'_{p}$  the branch $\omega_c(k)$ does this instead of
$\omega^\pi_P(k)$ .  Thus instability with respect to solutions
belonging to the family of solutions responsible for the long-wave
instability  turns into instability with respect to the pion
condensation.

Notice that the equation (\ref{6}) does not content the information
about the structure of the new ground state. It is possible that the
pion condensate is present in all cases. The experiments provide
  $g'_{NN}\sim 1$ \cite{AB1, EW}. Therefore
the instability which is interpreted as the appearance of the pion
condensation is related to $\omega_c(k)$ \cite{SR}.

The problem of the stability  of the Fermi liquid with the
pion exchange was discussed in papers \cite{BSJ, BBOW}. There the
$\pi$- and $\rho$-meson exchanges were considered as a source for
obtaining of the parameters of eq.(1). The meson exchange gives the
 spin-isospin and tensor contributions into the interaction (1).
Stability conditions put the restrictions on the combination of the
spin-isospin and tensor terms. However in the Migdal model \cite{AB1}
used here the pion exchange is taken into account in the direct
(annihilation) channel only.  Whereas the modification of the stability
conditions appears due to the tensor terms from the exchange diagrams
which are omitted in Migdal model.

In sect.4 the character of the instability related to the
$\omega_P(k)$ is investigated. Equation (\ref{5}) is symmetric with
respect to the substitution $\omega \leftrightarrow -\omega$ on the
physical sheet. Therefore  two branches of solutions differing by the
sign emerge simultaneously on the physical sheet. They are on the
positive and negative imaginary semiaxis. The both branches point out
to the instability of the ground state.  If the physical situation is
described by solution on the positive semiaxis then the amplitude of
the excitation $\omega_P(k)$ is increased as $\sim e^{|\omega_P(k)|t}$.
In the opposite case the instability is related
to the accumulation of the excitations with zero energy at nonzero
momenta $\omega_P(k\neq0)=0$. The question is, which of the solutions
describe the physical situation.

We suggest the answer to this question for the excitations with the
quantum numbers $J^\pi=0^-$.  Following \cite{AB3} we consider the
nuclear matter with one-pion exchange included.
 There are  rules which classify solutions on
pertaining to $\pi^+$- or $\pi^-$-type \cite{AB1,AB3} in this case.
We can obtain an answer by applying these rules to the all solutions of
(\ref{6}) (including $\omega^\pi_P(k)$) .  The result for the nuclear
matter without the inclusion of the pion exchange  can be obtained in
the limit $f_{\pi NN} \rightarrow 0$.  It is shown that the instability
of the ground state related to the breaking of the stability conditions
(\ref{3}) appears due to accumulations of the excitations with zero
energy $\omega_P(k\neq 0)=0$.

In paper \cite{Liu} the pion dispersion equation was study for the
relativistic pion propagator. The comparison of our results
 (here and in \cite{SR,DRS}) with the results of the paper \cite{Liu}
demonstrates the importance of the relativistic corrections.

\vspace{1cm}

\begin{center}
{\bf 1. The instability of the ground state with respect to the
long-wave excitations}
\end{center}

Let us calculate the particle-hole loop $A$  \cite{AB1,EW} and put the
obtained expression in (\ref{5}). The dispersion equation for
the frequency of the zero-sound is
\begin{equation} \label{9}
\frac{1}{f} = -4C_0[\Phi(\omega,k) +
\Phi(-\omega,k)],
\end{equation}
where $\Phi(\omega,k)$ stands for  excitation of the
particle-hole loop and $\Phi(-\omega,k)$ for absorption.
For  $0\leq k\leq 2p_F$  the function $\Phi(\omega,k)$ has the form
\begin{equation} \label{10}
\Phi(\omega,k) = \frac{m}k\
\frac1{4\pi^2}\left( \frac{-\omega m+kp_F}2-\omega
m\ln\left(\frac{ \omega m}{\omega m-kp_F+k^2/2}\right)\ +
\right.\nonumber
\end{equation}
$$ +\left.  \frac{(kp_F)^2-(\omega
m-k^2/2)^2}{2k^2}\ln\left( \frac{\omega m-kp_F-k^2/2}{\omega
m-kp_F+k^2/2}\right)\right).
$$
 while, $k > 2p_F$, it reduces to Migdal function
\begin{equation} \label{11}
\Phi(\omega,k) =\
\frac1{4\pi^2}\ \frac{m^3}{k^3}
\left[\frac{a^2-b^2}2\ln\left(\frac{a+b}{a-b}\right)-ab\right]\,
\end{equation}
where $a=\omega-k^2/2m$, $b=kp_F/m$.
However, the well-known expression of $\Phi(\omega,k)$ is the
Migdal function for all values of $k$. If substitute the expression
(\ref{10}) into (\ref{9}) and regroup terms it is easy to express
$\Phi(\omega,k)$ as the sum of two Migdal functions. But in this case
(\ref{9}) has two overlapping cuts for $0\leq k\leq 2p_F$. It is not
convenient for us since we study the branches of solutions on the
physical and unphysical sheets and the pass from one sheet to another
across the logarithmic cuts.  Therefore we use  (\ref{10}) for
$\Phi(\omega,k)$  at $k \leq 2p_F$  in present paper. Let us consider
the cuts of (\ref{9}). The cuts corresponding to the first and
second logarithms in (\ref{10}) we denote as $I$  and $II$.
The cuts of the function
$\Phi(-\omega,k)$ lie on the negative real semiaxis symmetrically with
respect to the cuts of $\Phi(\omega,k)$ (Fig.1).  Thus there are four
cuts for $k\leq2p_F$. There are two cuts  for $k > 2p_F$. They do not
overlap and do not tough the origin of coordinates.

\begin{figure}[h]%1
\centerline{\epsfig{file=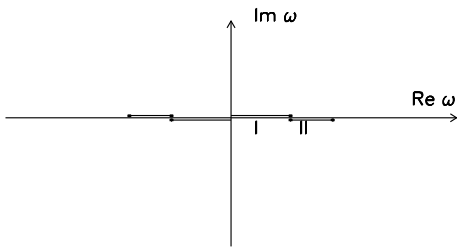,width=6cm}}
\caption{
 Cuts of eq.(\ref{9}) in the complex plane of pion frequency.
}
\end{figure}

\begin{figure}[h]%2
\centerline{\epsfig{file=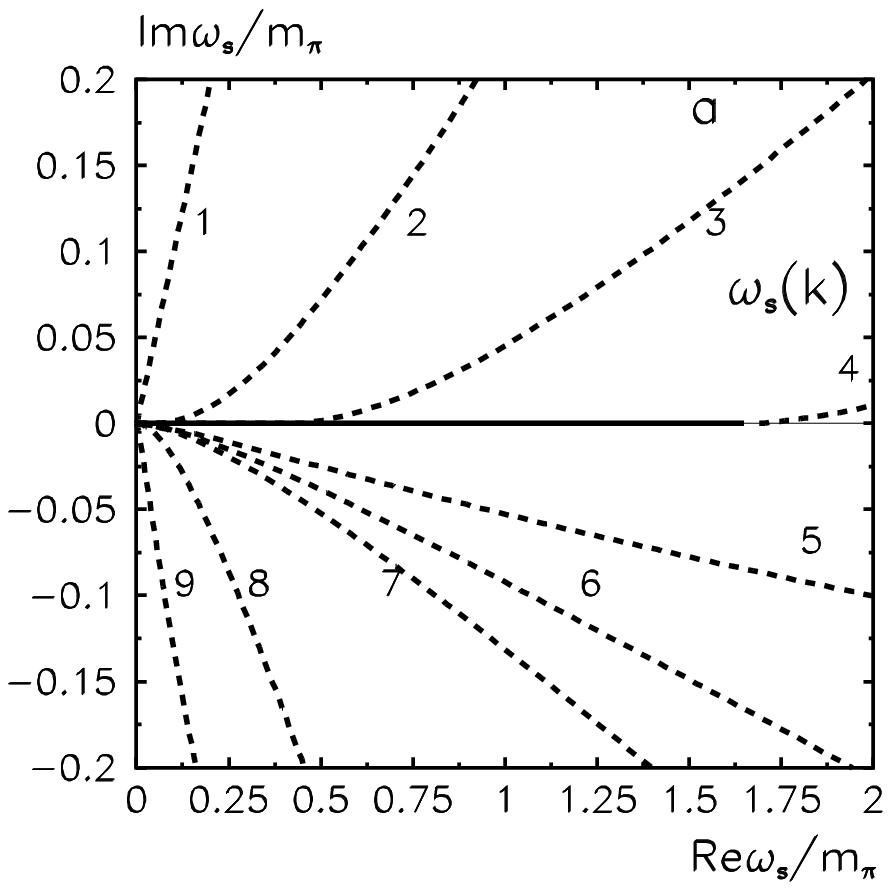,width=8cm}
\epsfig{file=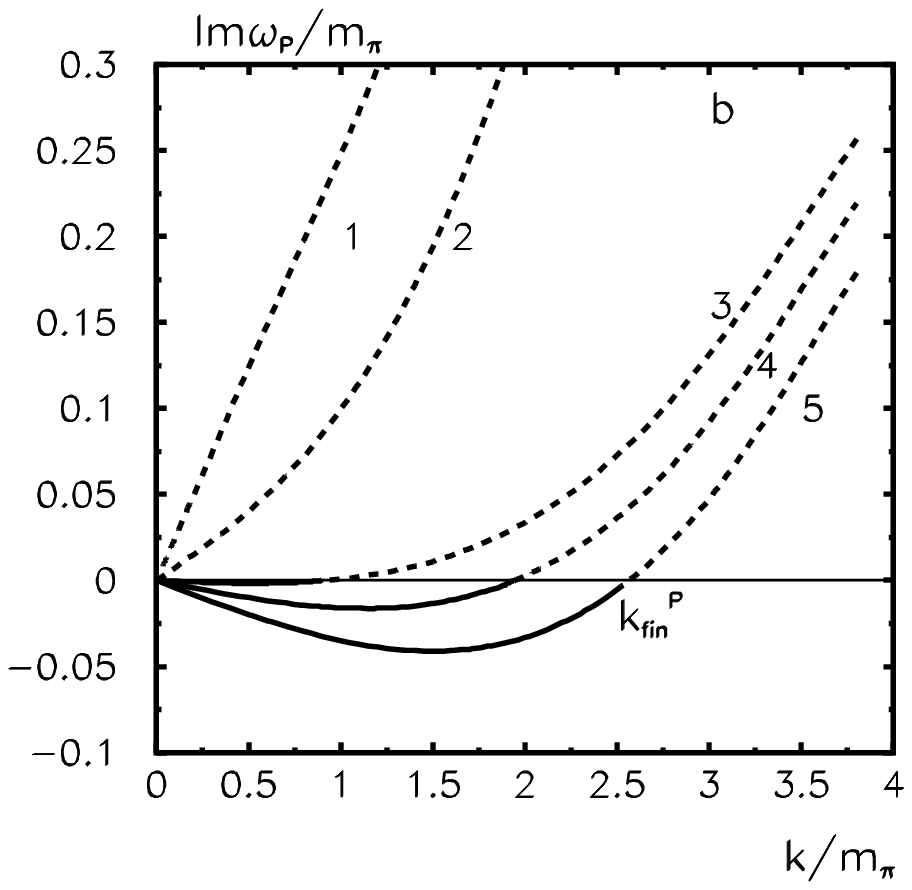,width=8cm}}
\caption{
a)~Branches $\omega_s(k)$ for the different values of $f$
in the complex plane of $\omega$.
The dashed parts of curves lie on the unphysical sheet,
 $\lambda=1/f$. The branches are drawn
for $\lambda$ = (1) 10$^4$, (2) 2.5, (3) 1.0, (4) 0.5, (5) -0.01,
(6) -0.5, (7) -1.0, (8) -10.0, (9) -10$^4$.
b)~Branches $\omega_P(k)$ for the different values of $f$.
Branches are presented for $f$=
 (1) -0.2, (2) -0.4, (3) -0.51, (4) -0.55, (5) -0.6.
}
\end{figure}

We note, that for $k/p_F\ll 1$ eq.(\ref{9}) reduce to the
well-known equation for the zero-sound frequency \cite{LP,PN}
$$
\frac1{f}=\frac{s}{2}\ln\frac{s+1}{s-1} - 1  $$
with $s=\frac{\omega m}{p_F k}$.
Turn now to the analysis of the solutions to (\ref{9}).
First consider  zero-sound branches $\omega_s(k)$.
We solve  Eq.(\ref{9}) for different values of $f$
both satisfying and not the stability condition
(\ref{3}). The results are shown in Fig.2a
\footnote{All variables on the figures are
given in the pion mass units ($m_\pi$=0.14 GeV). It is naturally for
the pion dispersion equation by is convenient for the zero-sound in
Fermi liquid also}.   For $f > 0$
the branch $\omega_s(k)$ lies on the real axis of the physical
sheet in the interval of momenta $k=(0,k^s_{fin})$
and has the values ranging from $\omega_s(k=0)=0$ to
$\omega_s(k^s_{fin})= v_Fk^s_{fin}$ ($v_F=p_F/m$ is the velocity of
quasiparticles on the Fermi surface \cite{PN}).

The branch $\omega_s(k)$ goes over under cut
$II$ to an unphysical sheet for $k>k^s_{fin}$.
The  parts of the branches  which lie on
the unphysical sheet (belonging to the Riemann surface of
the second logarithm in (\ref{10})) are dashed in Fig.2. Shifting $f$
to the negative values we obtain the solutions disposed in the low
semiplane of the same unphysical sheet.  In
Fig.2a we see that when   $\lambda=1/f$ decreases smoothly from large
positive ($10^4$, curve 1) to small values ($\lambda=0.5$ for curve 4
and $\lambda=-0.01$ for curve 5), and then to the large negative values
(-$10^4$, curve 9) the branches successively transform into one
 another. The branches $\omega_s(k)$ do not exhibit any singularity
that could result in a phase transition.

There is, however, yet another set of solutions of (\ref{9})
which leads to instability of the ground state.
We denote it as $\omega_P(k)$.
These are  pure imaginary solutions.
The branches  $\omega_P(k)$ lie on the
unphysical sheet (belonging to the Riemann surface of the
first logarithm in (\ref{10})), if the values of the parameters of the
effective interaction satisfy Eq.(2).
It is shown in Fig.2b that
branches $\omega_P(k)$ appear on the negative imaginary semiaxis of the
physical sheet for $f\leq -1/2$ at $k=0$.  They terminate under the cut
$I$ for $k=k^P_{fin}$.  We note that $k^P_{fin}$ increases while $|f|$
increases ($f<0$).  The equation (\ref{9}) is symmetric with respect to
the sign $\omega$ and  both solutions
$\omega_P(k)$ and $-\omega_P(k)$ are on the physical sheet.
This illustrates the fact that at $f\leq -1/2$ two imaginary solutions
appear \cite{T}.  Now consider the influence of the pion-nucleon
interaction on $\omega_s(k)$ and $\omega_P(k)$. We search the solutions
 $\omega^\pi_s(k)$ and $\omega^\pi_P(k)$ to the pion dispersion
equation (\ref{6}) corresponding to $\omega_s(k)$ and $\omega_P(k)$.
 Polarization operator $\Pi$ is written in a usual form
(without the isobar-hole contribution and scalar part) \cite{DFMM,DRS}
\begin{equation}\label{15} \Pi= \Pi^0_N/E, \quad E=1-\gamma\Pi^0_N/k^2,
\end{equation} $$ \Pi^0_N(k,\omega)=-4\left(\frac{f_{\pi
NN}}{m_\pi}\right)^2 [\Phi(\omega,k) + \Phi(-\omega,k)] $$ where $\Phi$
are given in (\ref{10}), (\ref{11}), and $\gamma=
C_0g'\left(\frac{m_\pi}{f_{\pi NN}}\right)^2$. It is easy to see that
condition $E=0$ is the same as Eq.(\ref{9}). The pion-nucleon vertex
contains form factor of pion $
d(k^2)=(\Lambda^2-m_\pi^2)/(\Lambda^2+k^2)$ ($\Lambda=0.667$ GeV).

Consider the branches $\omega^\pi_P(k)$. In Fig.3a we show
the changing of the branches $\omega^\pi_P(k)$ with increasing
 $f_{\pi NN}$. The curve 1 is the same as the curve 2 in Fig.2b
which is  $\omega_P(k)$ for $g'=-0.4$.
The behaviour of $\omega^\pi_P(k)$ is changed drastically from the
 physical point of view, when  $f_{\pi NN}$ increases.
For the experimental value $f_{\pi NN}=1$  the
imaginary $\omega^\pi_P(k)$ goes over to the physical sheet in interval
of the values of momenta $(k_1,k_2)$. This signals the presence of
instability.  It is seen that $k_1, k_2$ are not small in comparison
with $p_F$ and the springing up instability is not the long-wave one.

\begin{figure}[h]%3
\centerline{\epsfig{file=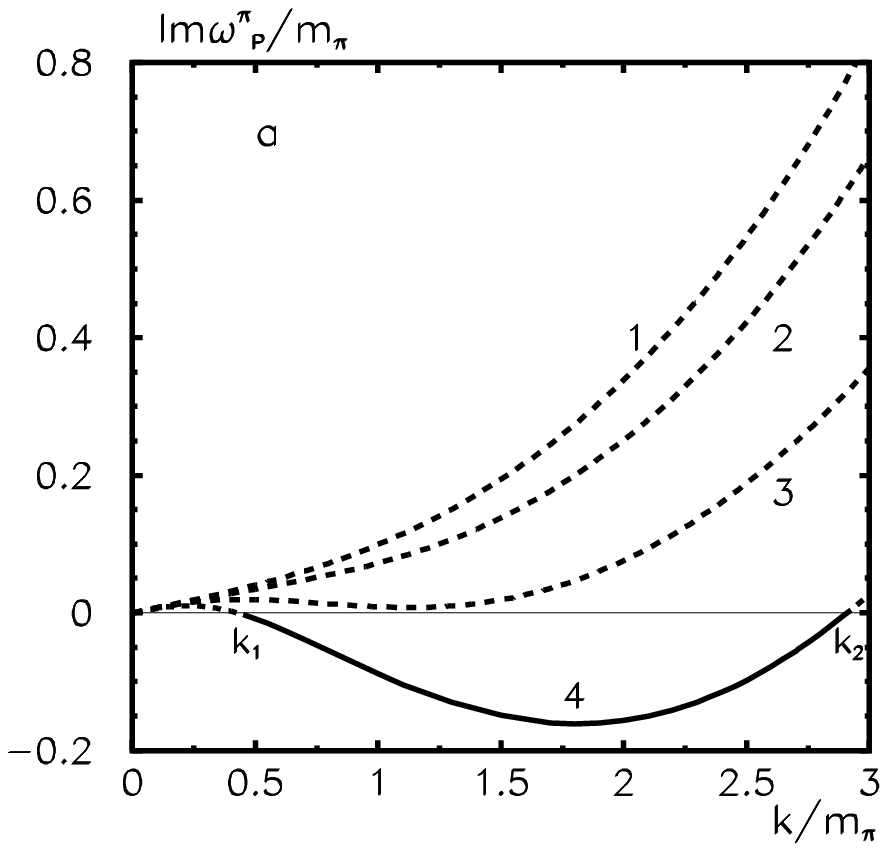,width=7cm}
\epsfig{file=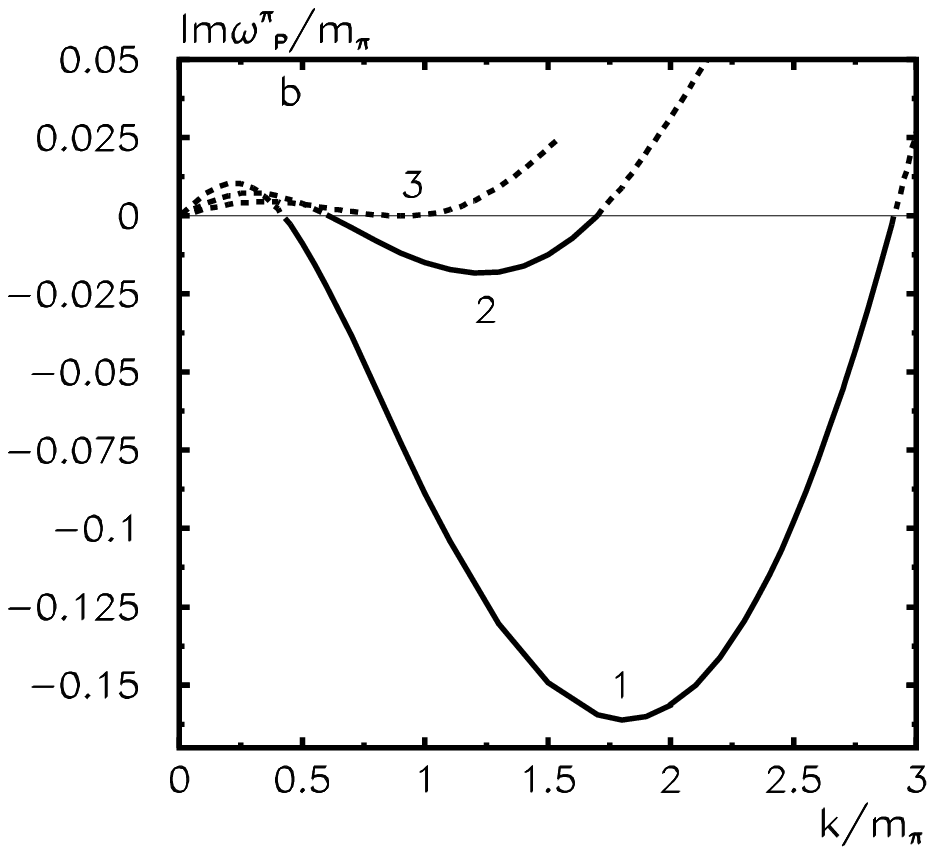,width=7cm}}
\centerline{\epsfig{file=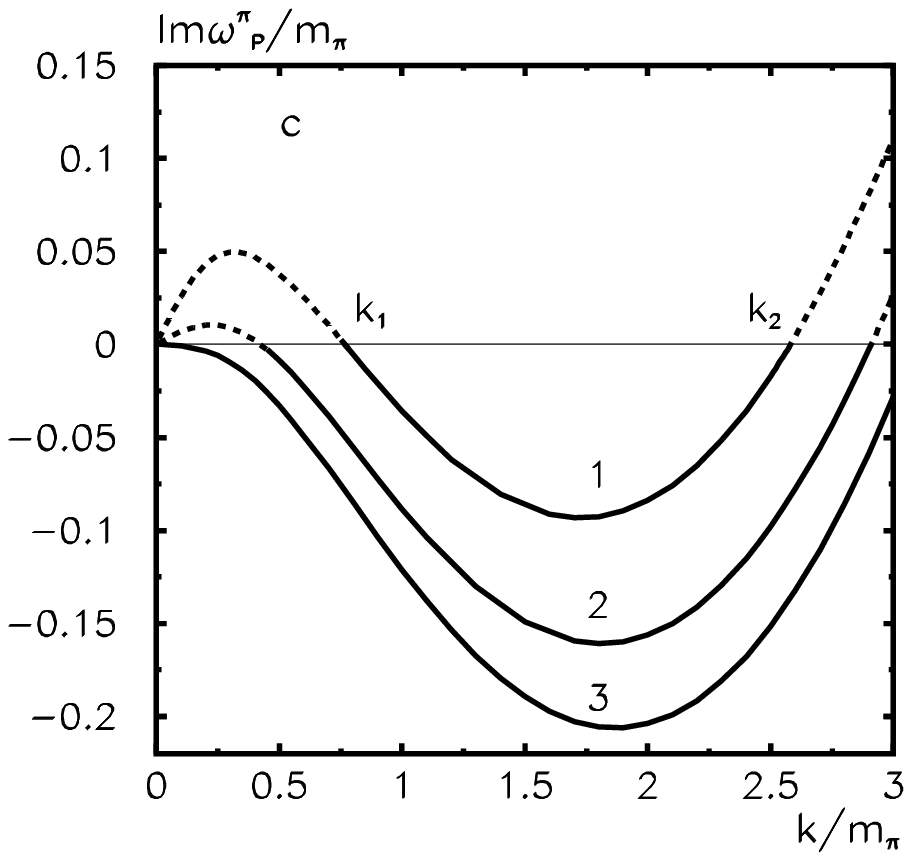,width=7cm}}
\caption{
The branches $\omega^\pi_P(k)$ .
The dashed parts of curves lie on the unphysical sheet.
a)~Dependence on pion-nucleon coupling constant is presented at
 $g'=-0.4$ and $\rho=\rho_0$ for: $f_{\pi NN}$ = (1) 0.0, (2) 0.3, (3)
 0.6, (4) 1.0.
b)~Dependence on the nuclear matter density at
 $g'=-0.4$ and $f_{\pi NN}$=1.0 for: $\rho/\rho_0$ = (1) 1.0, (2) 0.1.
c)~Dependence on the value of $g'$ at $\rho=\rho_0$ and $f_{\pi NN}$=1.0
  for: $g'$= (1) -0.2, (2) -0.4, (3) -0.55.
}
\end{figure}

The dependence of the instability on matter density is
demonstrated in Fig.3b. Curve 1 coincides with the curve 4 of Fig.3a
 and it is calculated for the equilibrium density  $\rho_0$. Curve 2 is
presented for the density $0.1\rho_0$.  The critical density for the
parameters $f_{\pi NN} = 1,$ $g'=-0.4$ is equal to $\rho_c=0.07\rho_0$.
For $\rho < \rho_c$ the nuclear matter is stable (curve 3).  In Fig.3c
the branches of solutions $\omega^\pi_P(k)$ for different $g'$ are
presented. We can see how the condition (\ref{3}) is realized by
$\omega^\pi_P(k)$.  For $g'$ larger then the critical one:  $0 > g' >
 -1/2$, the interval of the values of the momenta $k_1\leq k\leq k_2$
for which the solution is on the physical sheet decreases with
decreasing of the density (Fig.3b). The instability
disappears for $\rho < \rho_c$.  However, for $g' \leq -1/2$
the instability exists at any density. This is clear because the
branches $\omega_P(k)$  and $\omega^\pi_P(k)$ are close one to another
at small $k$ because of the weakness of the pion-nucleon interaction
(\ref{5.1}) and condition (2) does not depends on density.

\vspace{2cm}

\begin{center}
{\bf 2. The instability of the ground state with respect to the
pion condensation}
\end{center}

As it was shown by Migdal \cite{AB3} the pion dispersion equation
(\ref{6}) has the solutions  with  negative frequency squared.
Such solutions emerge if the density becomes larger that some critical
one.  This observation became the starting point of development of pion
condensation theory.

In Fig.4 \cite{SR} it is shown the behaviour of pion dispersion
 solutions in the complex plane of a pion frequency. The
solutions are presented for the pion dispersion equation with
inclusion of isobar-hole excitations and the scalar part of
polarization operator $\Pi_s$ \cite{DRS}. The values of parameters
are $g'_{NN}=1.0,$ $g'_{\pi \Delta}=0.2,$ $g'_{\Delta \Delta}=0.8$.
In Fig.4 the branches of solutions are presented for nonzero isobar
width  $\Gamma_\Delta=115$ MeV. Owing to the width it is easier to
trace the behaviour of the branches as functions of $k$. When
$\Gamma_\Delta=0$  the branches $\omega_\pi$, $\omega^\pi_s$ and
$\omega_\Delta$ become real and $\omega_c$ pure imaginary
(Im~$\omega_c(k)\leq 0$) (Fig.4) on the physical sheet.  It is
$\omega_c$ that has the negative frequency squared. The critical
density in these calculations is $\rho_c\simeq 1.2\rho_0$, this
correspond to Fermi momentum $p_{Fc}=$283 MeV.  In Fig.4 the solutions
are presented for the values of Fermi momenta which are a little
smaller than $p_{Fc}$, i.e. $p_F$=268 MeV (corresponding to equilibrium
density) and a little larger than $p_{Fc}$, i.e. $p_F$=290 MeV.
The latter  case demonstrates
the appearance of $\omega_c(k)$ on the physical sheet).

\begin{figure}[h]%4
\centerline{\epsfig{file=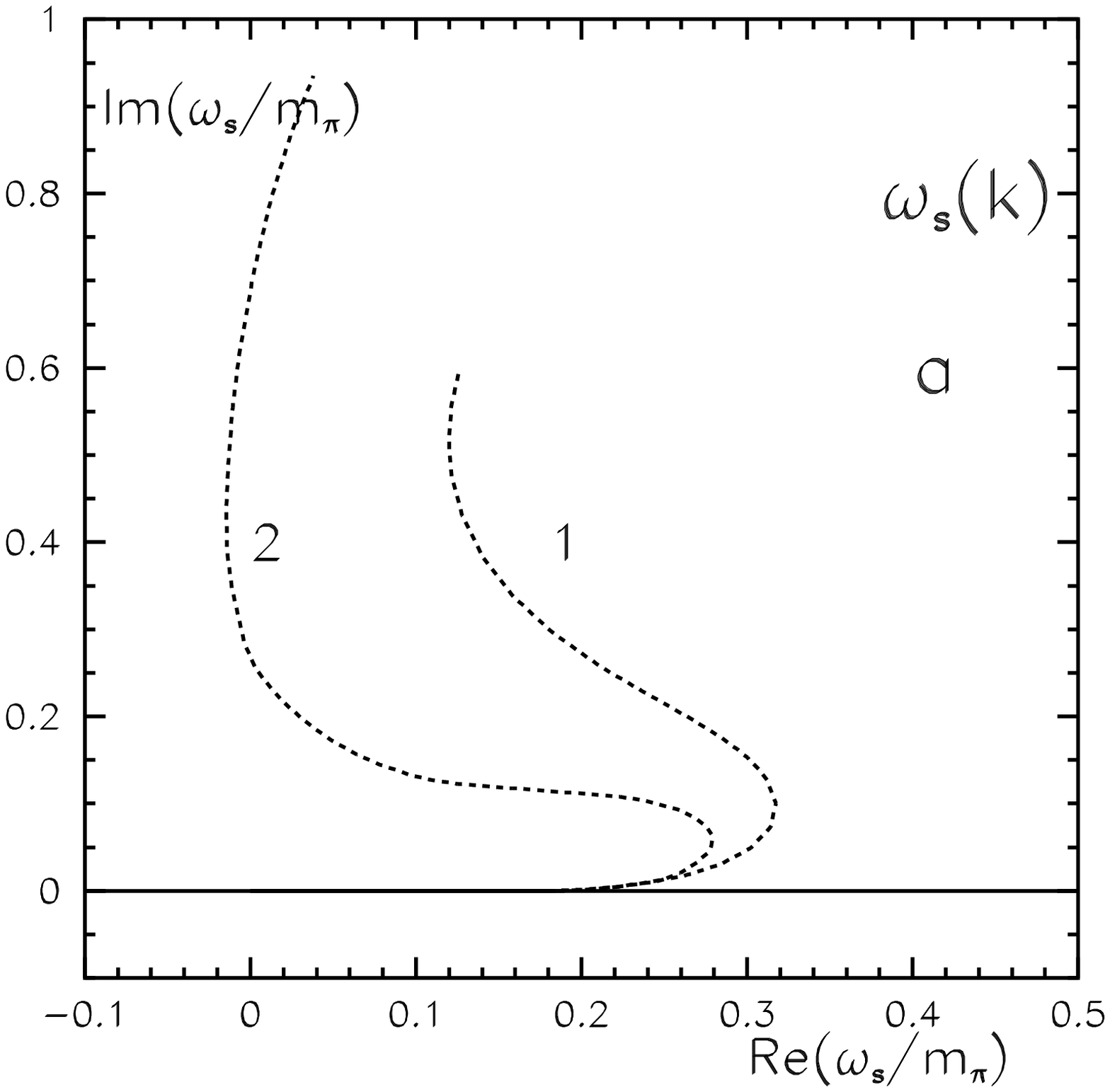,width=6cm}
\epsfig{file=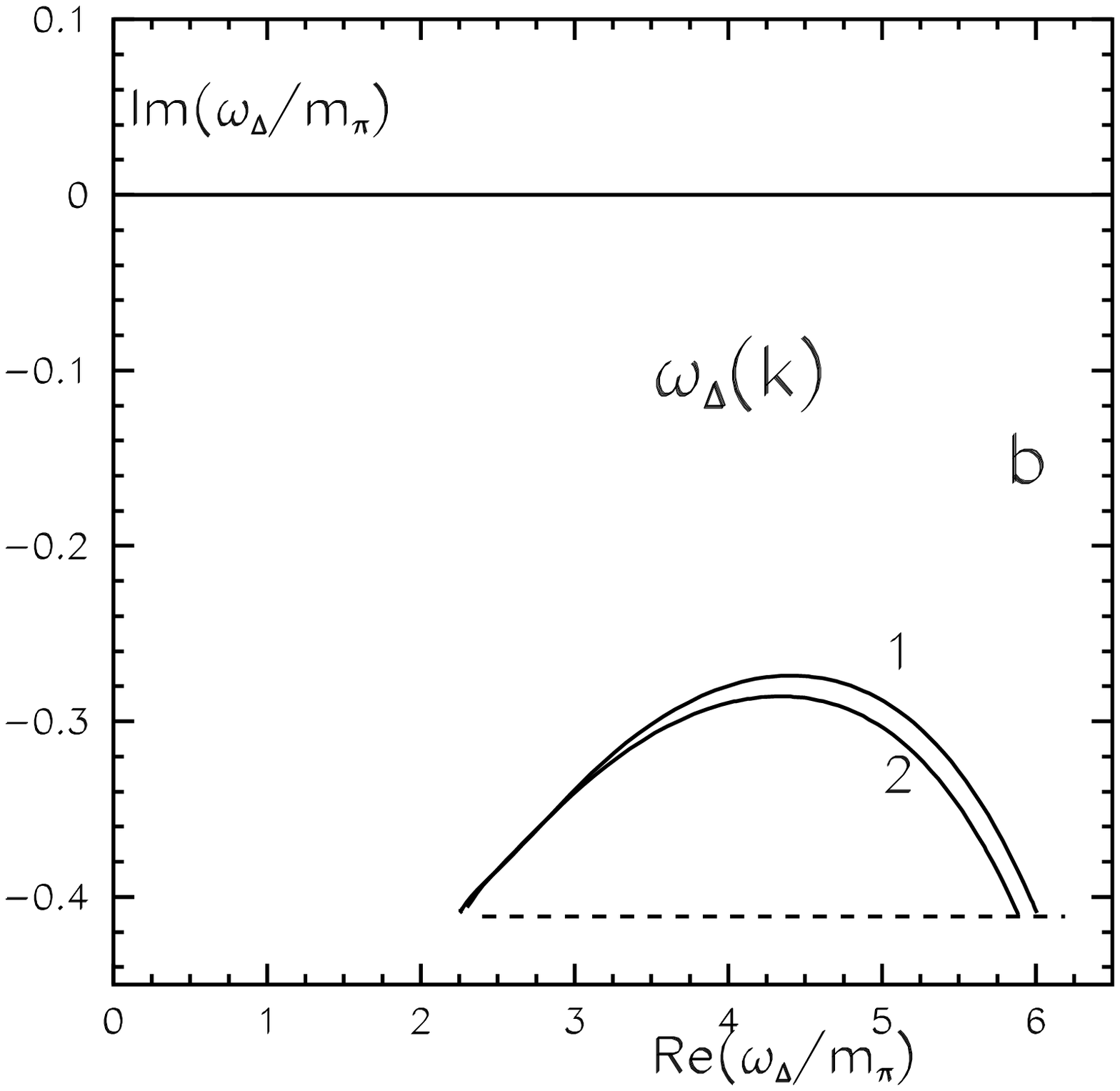,width=6cm}}
\centerline{\epsfig{file=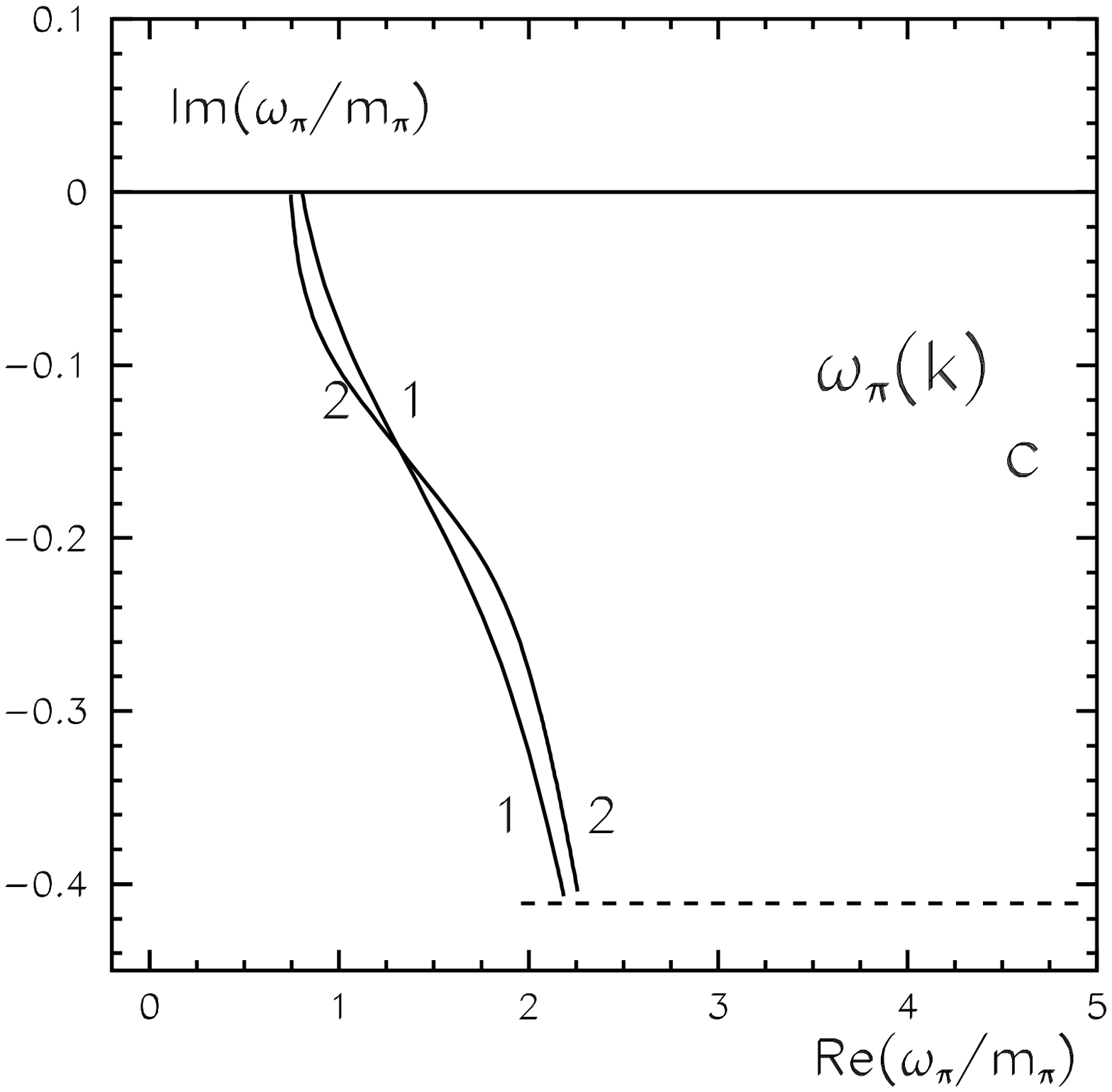,width=6cm}
\epsfig{file=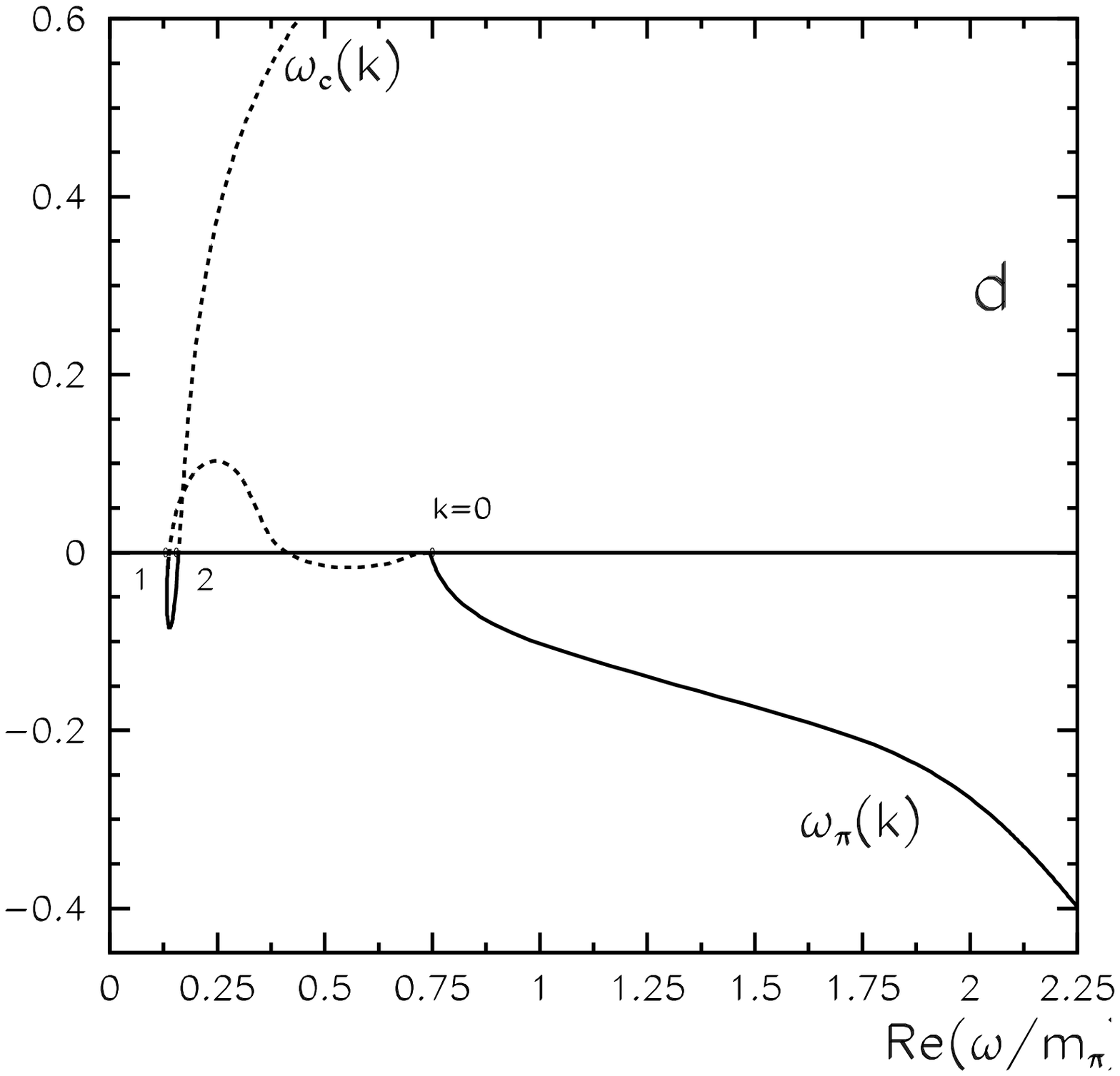,width=6cm}}
\caption{
Branches of  solutions to Eq.(\ref{6}) in the complex plane
  of $\omega$:
a)~zero-sound wave $\omega^\pi_s(k)$,
b)~isobar branch $\omega_\Delta(k)$ [the horizontal dashed line
  depicts the cut of isobar polarization operator \cite{SR} at $p_F$ =
  290 MeV for the momentum value at which $\omega_\Delta(k)$  goes
  under the cut],
c)~pion branch $\omega_\pi(k)$ [the horizontal dashed line has the
  same meaning as in Fig.(b)],
d)~the branches $\omega_\pi(k)$ and $\omega_c(k)$ at $p_F$=290 MeV.
The curves 1 (2) corresponds to $p_F$= 268
(290) MeV.  The dashed parts of curves lie on the unphysical sheet.  }
\end{figure}

In Fig.4a  the zero-sound branch  $\omega^\pi_s(k)$ is shown.
It begins at the point $\omega^\pi_s(k=0)=0$, moves along the real axis
 and goes over under the cut $II$ at  $k^s_{fin}$.  In Fig.4b the
isobar branch $\omega_\Delta(k)$ is shown, it issues from the point
$\omega_\Delta(k=0)=m_\Delta - m_N$ and terminates under the cut of the
isobar polarization operator \cite{DRS}. In Fig.4c the pion branch
$\omega_\pi(k)$ is presented, $\omega_\pi(k=0)=m^*_\pi$, $m^{*2}_\pi =
m^2_\pi + \Pi_s.$ It terminates under the isobar cut as well.  In
 Fig.4d the relative placement of $\omega_\pi(k)$ and $\omega_c(k)$ is
presented. The branch $\omega_c(k)$ appears on the physical sheet and
 returns on the same unphysical sheet under the cut $I$.

The dependence of  $\omega_c(k)$ on the isobar width and nuclear
density is demonstrated in Fig.5. In Fig.5a the behaviour
of $\omega_c(k)$ at $\Gamma_\Delta\rightarrow 0$ is shown . The curve 1
corresponding to
$\Gamma_\Delta=0$  shows that the solutions on the physical sheet
are pure imaginary. In Fig.5b we show the dependence of
 $\omega_c(k)$ on density. When the Fermi momentum is less than the
critical one $p_F < p_{Fc}$  the branch does not appear on the
physical sheet  (curve 1) and lies on the unphysical sheet
completely.

\begin{figure}[h]%5
\centerline{\epsfig{file=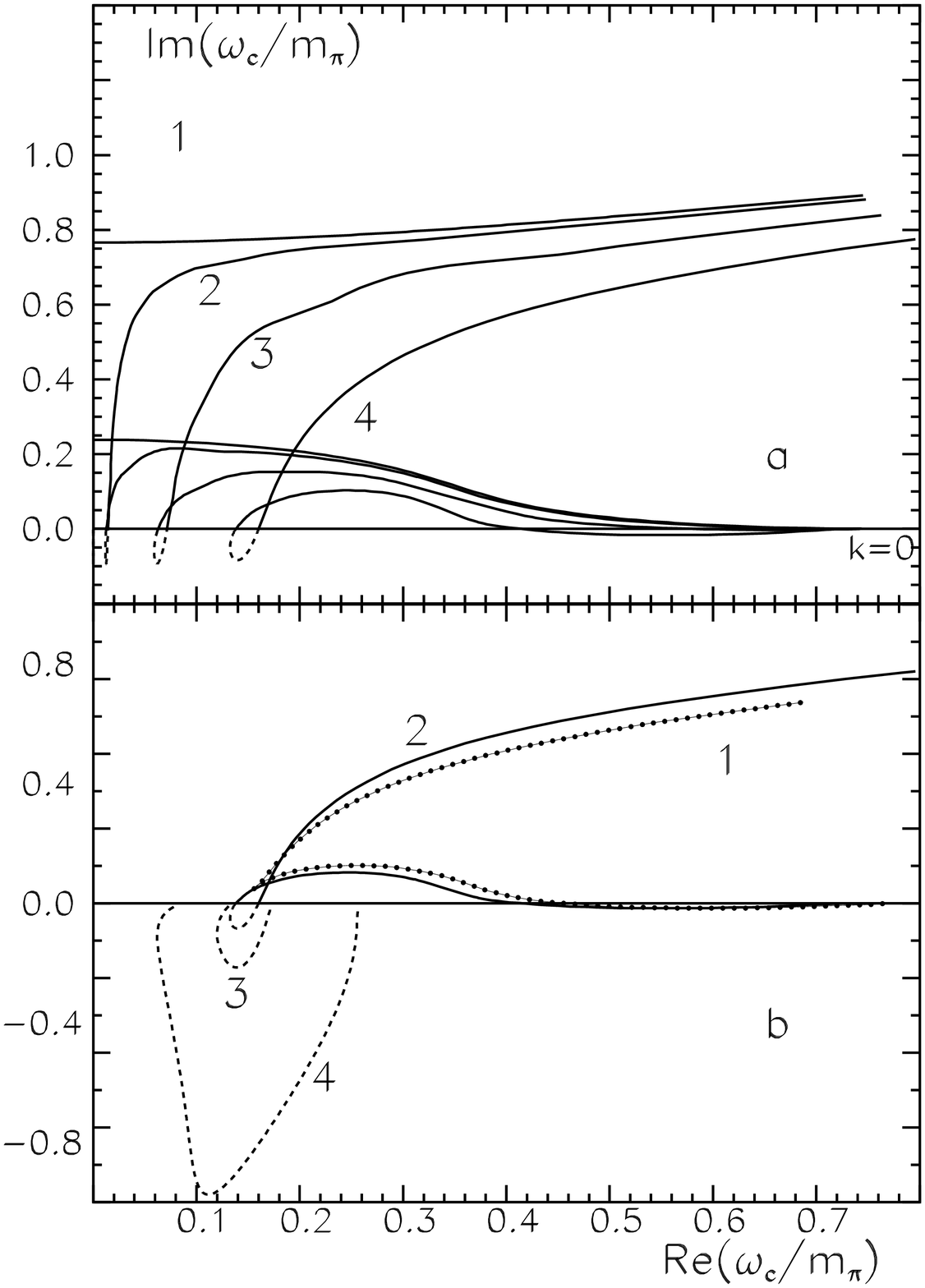,width=8cm}}
\caption{
Condensate branch  $\omega_c$ in the complex plane of
$\omega$ [the dashed (solid) curves represent those parts of the
solutions branches that lie on the physical (unphysical) sheets]:
a)~branch $\omega_c(k)$ at $p_F$=290 MeV for isobar-width values of
$\Gamma_\Delta$ = (1) 0, (2) 10, (3) 50 and (4) 115 MeV; b)~branch
$\omega_c(k)$ at $\Gamma_\Delta$ = 115 MeV for Fermi momentum
values of $p_F$ = (1) 280, (2) 290, (3) 300 and (4) 360 MeV.
 Only the solutions on the physical sheet are displayed for the
curves 3 and 4. Curve 1 for $p_F$=280 MeV completely lies on unphysical
sheet.
}
\end{figure}

For the further analysis of the interaction of branches $\omega_c(k)$
and $\omega_P(k)$  we need more detailed picture of the
behaviour of the solutions on the unphysical sheet. In Fig.6 we use the
same model as in Figs.4,5 with $\Gamma_\Delta=0$ \cite{DRS}.
The parts of branches which lie on the imaginary axis
are shifted in different sides to demonstrate the change of branches
with $k$.  In Fig.6  a new branch  $\omega^1_c(k)$ is presented besides
$\omega_c(k)$. It does not appear on the physical sheet but it is the
third ingredient of the interaction of $\omega_c(k)$ and
$\omega^\pi_P(k)$.  We see that
the branch $\omega_c(k)$ and symmetric branch $\omega^1_c(k)$ coincide
in the imaginary axis (at $k_A=1.09 m_\pi$) on the unphysical sheet in
point $A$. At larger $k$ the branch  $\omega_c(k)$ goes down along the
imaginary axis and goes over to the physical sheet. The branch
$\omega^1_c(k)$ goes up. At $k\simeq 1.8 m_\pi$ the branches turn back
and meet one another again in the point $B$ ($k_B=3.08 m_\pi$) then
they pass in the opposite sides on the unphysical sheet of complex
plane. Fig.6 clarifies the behaviour of the curve 1 in Fig.5a when
function  sharply changes the dependence on $k$ in two points.  The
branch $\omega_c(k)$ is pure imaginary when the value of $k$ changes in
the interval between $k_A$ and $k_B$.

\begin{figure}[h]%6
\centerline{\epsfig{file=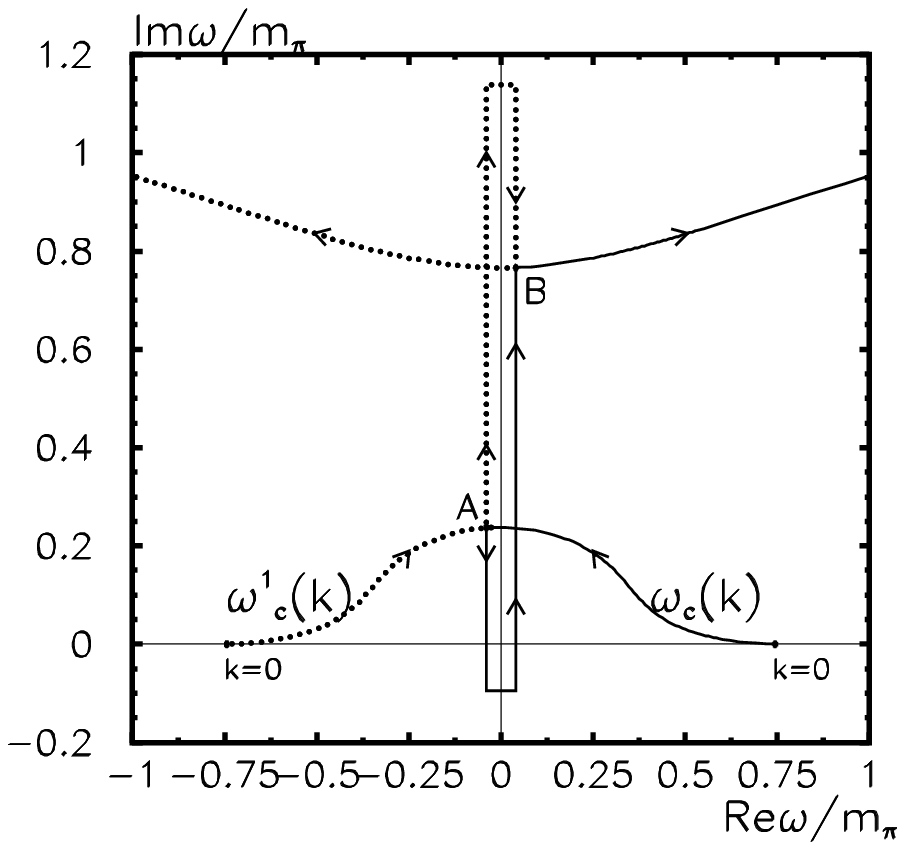,width=8cm}} \caption{ Branches
$\omega_c(k)$ and  $\omega^1_c(k)$ in the complex plane of $\omega$,
$p_F$ = 290 MeV.  The part of the complex plane above the x-coordinate
axis is the unphysical sheet and the semiplane below the x-axis is a
physical sheet.  The arrows show the direction of increasing of the
momentum $k$ along the branches.
}
\end{figure}

Note that at the critical density $\rho_c$ the branch
$\omega_c(k)$ appears on the physical sheet at the only point
 $k=k_c$.  At this point $\omega^2_c(k_c)=0$.  This solution is a
signal that the phase transition in nuclear matter takes place
\cite{AB1}.  The solutions at larger densities are not
related to a real physical object. Nevertheless, it is very useful to
consider the solutions at $\rho \geq \rho_c$.

\vspace{1cm}
\begin{center}
{\bf 3. Interaction of branches $\omega_P^\pi(k)$ and $\omega_c(k)$
}
\end{center}

Let us consider the solutions to the pion dispersion equation
on the physical sheet for $g'$ near zero $g'=(-0.05,0.05)$.
[Now we return to the version of calculation without the isobar and
scalar part of polarization operator ($m^*_\pi=m_\pi$).]
In Fig.7 we show how the branches change  with $g'$.
The branches  $\omega_c(k)$ is shown by dashed curves while
$\omega^\pi_P(k)$ is presented by the solid line. Recall that the
branches go over to the physical sheet across the cut $I$. Looking at
Fig.7a we can not notice the transition from one type of instability to
another. But this transition becomes visible if we turn to Fig.7b,
where the same results are presented in the larger interval of momenta.
The curves 4, 5 start at the point $\omega=0$ at $k=0$ whereas curves
1, 2, 3 are drawn from the momentum $k=k_A$ only, since they are
complex at $k<k_A$ (compare with Fig.6 made in the complex plane). They
behave themselves as $\omega_c(k)$ in Fig.6, starting from
$\omega=m_\pi$ at $k=0$.

\begin{figure}[h]%7
\centerline{\epsfig{file=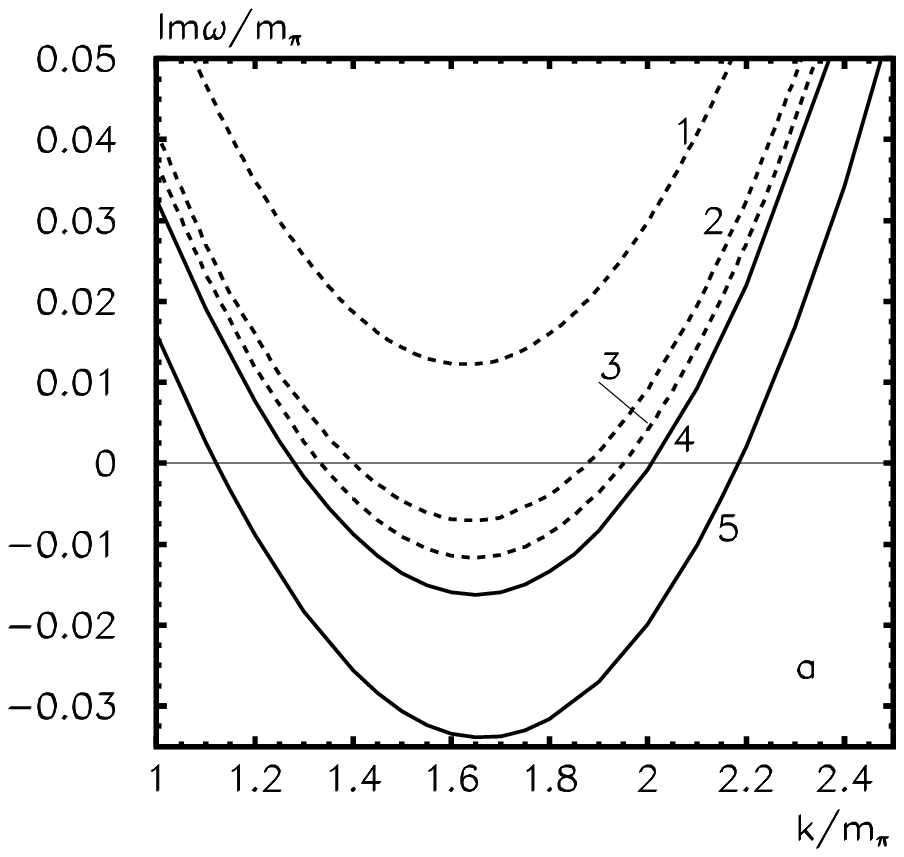,width=8cm}
\epsfig{file=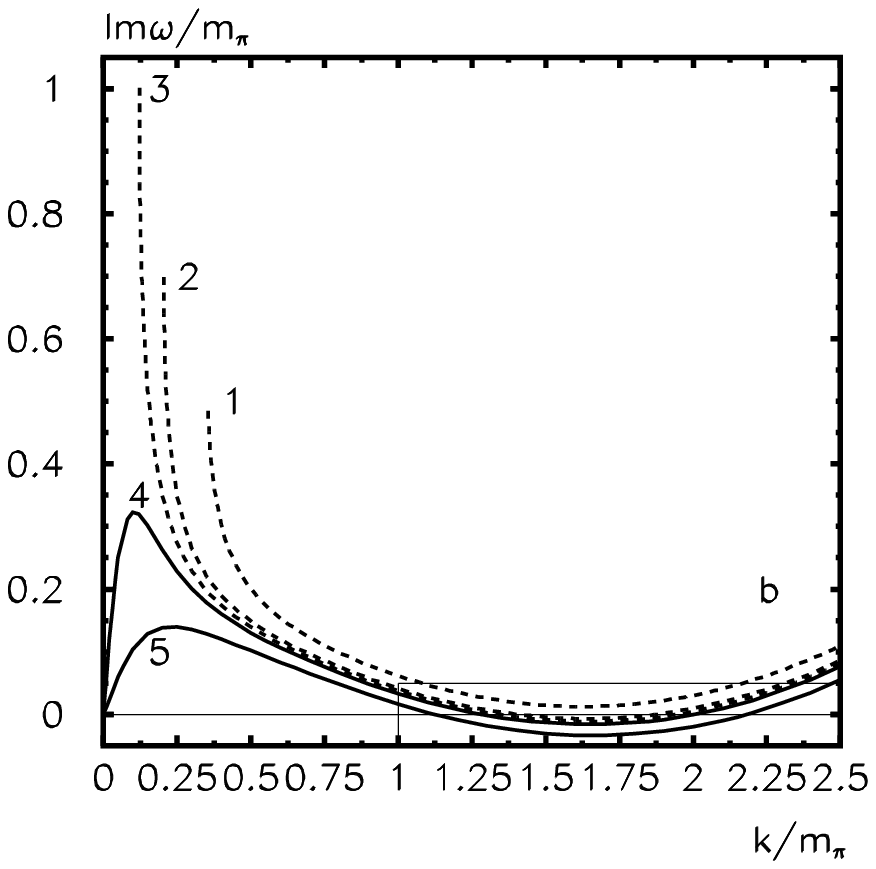,width=8cm}}
\caption{
Imaginary solutions of (\ref{10}) for $g'$= (1) 0.05, (2)
0.01, (3) 0.0, (4) -0.01, (5) -0.05.
The part of the complex plane above the
x-coordinate axis is the unphysical sheet and the semiplane below the
x-axis is a physical sheet.
The solid (dashed) curves correspond to  $\omega_P(k)$
($\omega_c(k)$).  a)~The interval of momentum values  $k/m_\pi$ = (1.0,
2.5); b)~the interval of momentum values  $k/m_\pi$ = (0.0,2.5). The
rectangle on the right below is the figure a).
}
\end{figure}

As mentioned above  the appearance of
$\omega^\pi_P(k)$ or $\omega_c(k)$ on the physical sheet takes
place at the density larger the critical $\rho \geq \rho_c$ at $g' >
-1/2$. The value $\rho_c$ is determined by parameters of matter:  $m$,
$g'$, $f_{\pi NN}$.

We can not determine, if the replacement of the branches took place if
we limit ourselves to the physical sheet only. This is illustrated in
Fig.8.
Here the dependence of the critical Fermi momenta $p_{Fc}$ on $g'$ is
demonstrated (i.e. the values of $p_F$ for which (\ref{6}) has
solutions $\omega=0$ for different $g'$ at the momenta $k=k_c$).
Looking at the curve we cannot say that the instability with respect to
$\omega_P(k)$ is changed by the instability with respect to
$\omega_c(k)$.  However the investigations similar to those,
demonstrated in Fig.7, show that the replacement takes place at
$g'=-0.0009$. The dependence of $k_c$ on $g'$ (dashed curve)
demonstrates how the instability becomes the long-wave one if
$g'\rightarrow -1/2$.

\begin{figure}[h]%8
\centerline{\epsfig{file=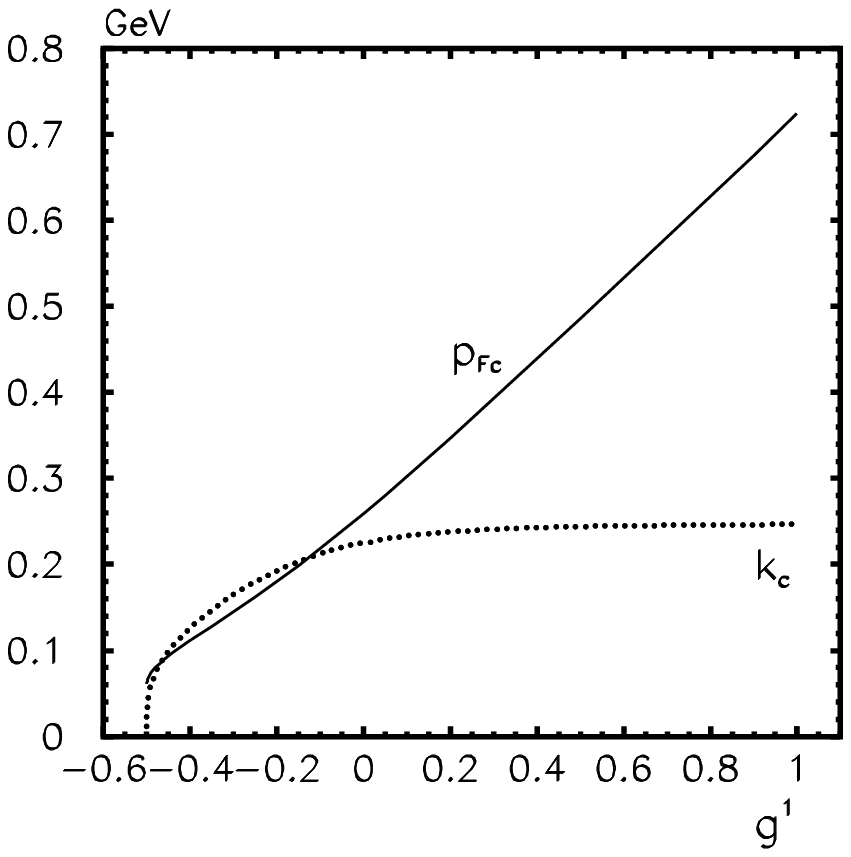,width=8cm}}
\caption{
Dependence of the critical Fermi momentum $p_{Fc}$ on
$g'$ (solid line). Dependence of $k_c$ on $g'$ (dashed line).
}
\end{figure}

A mechanism of the replacement of branches is presented in Fig.9.  We
show three interacting branches: $\omega_c(k)$, $\omega^1_c(k)$ and
$\omega_P(k)$ at values $g'$ near $g'_{p}$ in the complex plane of
$\omega$.  In the present calculation we have -0.00102 $< g'_{p} <
 $-0.001. In Fig.9a  at $g'=-0.00102$ the branch $\omega^\pi_P(k)$
moves between $\omega_c(k)$ and $\omega^1_c(k)$: $k_3=0.08 m_\pi,$ and
$k_2=0.091 m_\pi$. At $k > k_3$ $\omega^\pi_P(k)$ goes down and
penetrates to the physical sheet (we can compare with Fig.3). The
minimal value of $\omega^\pi_P(k)$ is reached at $k_{min}=1.65 m_\pi$.

\begin{figure}[h]%9
\centerline{\epsfig{file=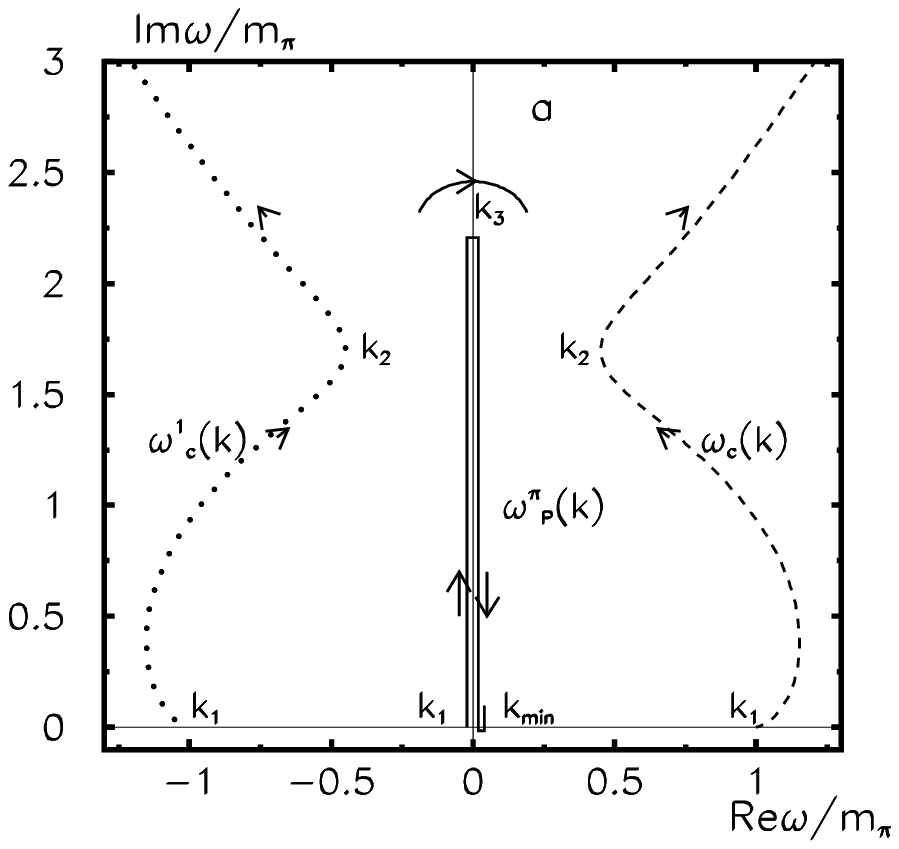,width=8cm}
\epsfig{file=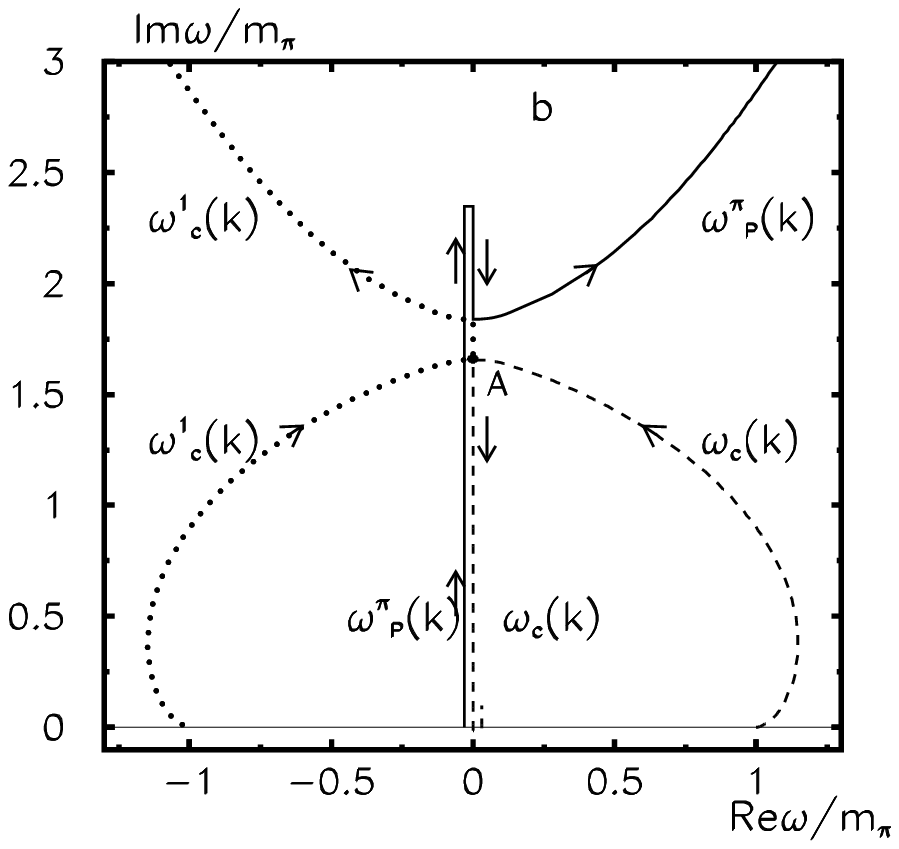,width=8cm}}
\caption{
Interaction of the branches $\omega_P(k)$ (solid),
$\omega_c(k)$ (dashed) and $\omega^1_c(k)$ (dotted): a)~$g'$
  = -0.00102,  $k_2/m_\pi=0.091$, $k_3/m_\pi=0.08$,
$k_{min}/m_\pi=1.65$; $\omega_P(k)$
goes over to the physical
sheet, b)~$g'$=-0.001, $k_A/m_\pi$=0.09238;
 $\omega_c(k)$ goes over to the physical sheet.
The part of the complex plane of $\omega$ above the x-coordinate
axis is the unphysical sheet, below the x-axis is a physical sheet.
Momentum $k_1$=0.
The arrows show the direction of increasing of the
momentum $k$ along the branches.
}
\end{figure}

In Fig.9b at $g'=-0.001$  the branches $\omega_c(k)$ and
 $\omega^1_c(k)$ blockade $\omega^\pi_P(k)$.  The branches
$\omega_c(k)$ and  $\omega^1_c(k)$ meet  in the point
 $A$ on the imaginary axis ($k_A=0.09238 m_\pi$).
The branch $\omega^1_c(k)$ goes up and meets the branch
$\omega^\pi_P(k)$  at $k=0.09242 m_\pi$ . After that  these branches
go to the opposite sides on the complex plane. But  at $k > k_A$
the branch $\omega_c(k)$  goes down  to the physical
sheet, as it was discussed in Fig.6 in detail.

Thus, there is a replacement  of branches which go over to physical
 sheet and responsible for the instability of the ground state.
The branch $\omega^\pi_P(k)$ is changed by $\omega_c(k)$.
It was explained in the previous section that $\omega^\pi_P(k)$
belongs to the family of solutions, which gives the long-wave
instability at $k\rightarrow 0$ and $g'\rightarrow-1/2$. On the
other side,  the instability related to $\omega_c(k)$ is interpreted
 as appearance of the pion condensation.

\begin{center}

{\bf 4. Investigation of the character of instability of the
ground state related to $\omega_P(k)$ and $\omega^\pi_P(k)$}

\end{center}

The existence of the solution of the dispersion equation with
$\omega_i^2(k) \leq 0$ on the physical sheet of the complex plane of
$\omega$ demonstrate the instability of the ground state. The equations
(\ref{6}), (\ref{9}), (\ref{15}) are symmetric with respect to the sign
of $\omega$ on the physical sheet, therefore we have two solutions
differing by sign.  In Figs.2b,3,6,7 there are presented solutions
which are placed on the negative imaginary  semiaxis on the physical
sheet.  Besides there are solutions on the positive semiaxis.

In this section we investigate which of solution describes the
instability of the ground state. Turn to the pion dispersion equation
(\ref{6}).  The following selection rules for physical solutions has
 been formulated by Migdal \cite{AB4}.
 The rules sort the solutions on relating to $\pi^+$- or $\pi^-$-type
 \cite{AB1,AB3}.  All solutions with
 \begin{equation} \label{16} \frac
 d{d\omega}(\omega^2-m^2_\pi-k^2-\Pi) > 0 \end{equation}
 at
$\omega=\omega_i(k)$, correspond to $\pi^+$-meson type, while those
 with
 \begin{equation} \label{17}
\frac d{d\omega}(\omega^2-m^2_\pi-k^2-\Pi) < 0
\end{equation}
 at $\omega=\omega_i(k)$ after the substitution
$\omega=-\omega_i(k)$ give the $\pi^-$-meson dispersion relation.
For $\pi^+$ type excitation we compare
the values $\varepsilon_k=\omega_i(k)$ with the physical energies
while for $\pi^-$ type we compare the values with the opposite sign
 $\varepsilon_k=-\omega_i(k)$ with the physical energies .  These rules
are written for the real solutions  $\omega_i(k)$ and we cannot apply
them to  $\omega^\pi_P(k)$ presented above.

However, the behaviour of $\omega_P^\pi(k)$ and $\omega_P(k)$ is
 changed if we take a large $|g'|$ ($g' < 0$).  In Fig.10a
the branch $\omega_P(k)$  at $g'=-2$ is presented.  We see that at the
large negative values of $g'$ (or $f, f', g$)  part of branch lies on
the real axis. The explanation  is the following. In Fig.2b we have
seen that the value of  $k_{fin}^P$ increases simultaneously with
$|g'|$.  The momentum $k_{fin}^P$ denotes the final momentum when the
branch leaves the physical sheet. It occurs that at $g'\simeq -2$ the
value $k^P_{fin} > 2p_F$. It was mentioned above that the cuts of the
polarization operator at $k > 2p_F$ do not tough the origin of
coordinates. Thus, the branch $\omega_P(k)$ (Fig.10a) returns to the
origin of coordinates along the imaginary axis having the momentum
$k_{int} > 2p_F$ (point 2), but there is no the cut.  As it is shown
in Fig.10a the branch turns to the real axis at $k>k_{int}$. While
$k$ increases, this branch penetrates through the cut
(at $k=k^P_{fin},$ point 3) and goes to the unphysical sheet under the
cut.  In some interval of momenta there is a real solution .  In Fig.10b
the analogous picture for $\omega^\pi_P(k)$ is presented (curve 1,
$g'=-3$).   The calculations in Fig.10b are carried out with isobar-hole
excitations and the scalar part of the polarization operator being
taken into account \cite{DRS}.

\begin{figure}[h]%10
\centerline{\epsfig{file=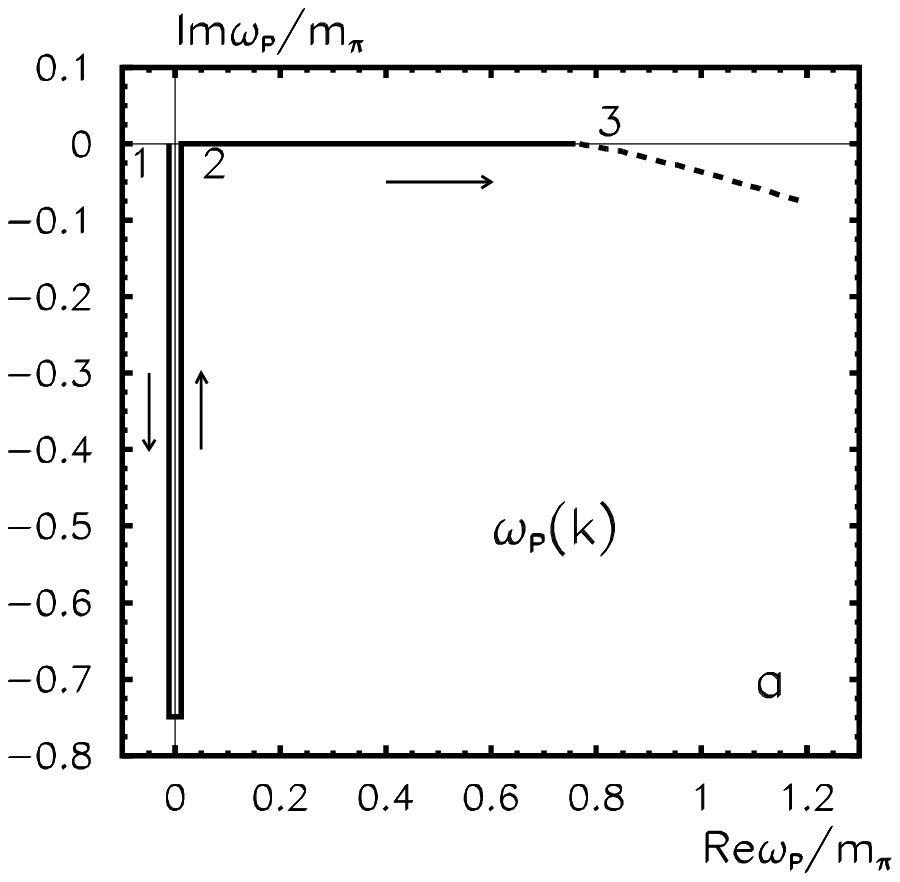,width=8cm}
\epsfig{file=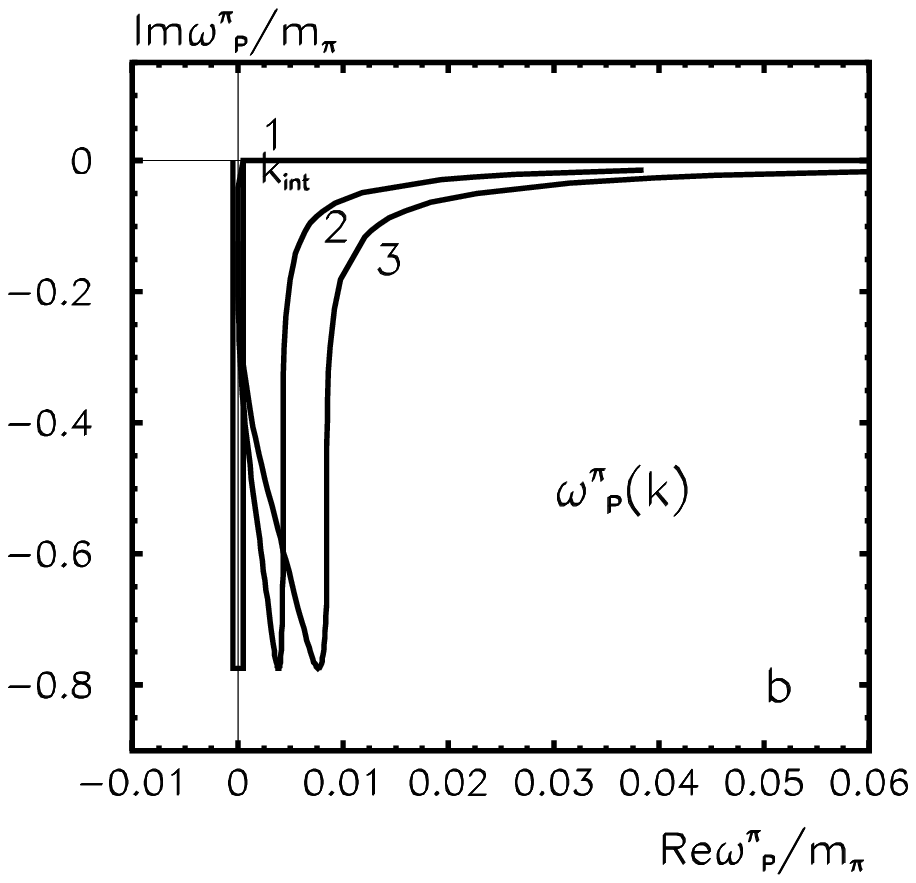,width=8cm}}
\caption{
a)~The branch
$\omega_P(k)$ at $g'=-2.0$ on the physical sheet of the complex plane
of $\omega$. The beginning of branch (point 1): $k_1$=0; branch turns
from imaginary to real meanings in (2): $k_2/m_\pi=k_{int}$=4.8; it
terminates under the cut (3)  $k_3/m_\pi=k^P_{fin}$=5.4.  b)~The branch
$\omega^\pi_P(k)$ at $g'$=-3.0.  The full pion polarization operator
 \cite {DRS,SR} is used.  The dependence of $\omega^\pi_P(k)$ on
isobar-width $\Gamma_\Delta$ = (1) 0, (2) 50, (3) 100 MeV.
}
\end{figure}

Now we apply the rule (\ref{16}) to the part of $\omega^\pi_P(k)$
which is purely real (curve 1, Fig.10b) and obtain that this
part of branch belongs to $\pi^+$-type. But we cannot find
 whether
the continuation of the real part to smaller $k < k_{int}$ lies on
the positive or negative imaginary semiaxis. This may be checked
easily including isobar width into consideration. The curves 2 and
3 in Fig.10b correspond to $\Gamma_\Delta$= 50 and 100 MeV. At
 $\Gamma_\Delta\rightarrow 0$, the curves approach the curve 1.

Now we can draw a conclusion that curve 1 (Fig.10b)
 lying on the  negative imaginary semiaxis at $k < k_{int}$
belongs to $\pi^+$-type excitations.
This conclusion is the only information we need. Considering the
pion dispersion equation at large  $|g'|$ ($g' < 0$) we remember
 that there is a phase transition at $g'\leq -1/2$ at the any density.
Then, the values of $k$ are too large for the simple model of
zero-sound. But it was demonstrated above that there is a continuous
passage on $g'$ and $k$ to the range of values where the Fermi liquid
 theory  is valid.

In correspondence with (\ref{16}) the physical energies which
determine the character of the instability are
$\varepsilon_k=\omega_P(k)=-i|\omega_P(k)|$. It means that the
instability emerges due to accumulation  of the excitations with
zero energy in the ground state but not because of the exponential
 increasing of  solution amplitudes.

As it is shown in Fig.3a (and can be easily seen from
(\ref{6})) there is a continuous passage from $\omega_P^\pi(k)$ to
 $\omega_P(k)$ when $f_{\pi NN}\rightarrow 0$. Therefore the
 conclusion  about the character of instability remains true for
 $J^\pi = 0^-$ wave for the nuclear matter without pion-nucleon
 interaction taking into account as well.

\begin{center}

{\bf  Summary}

\end{center}
In this paper the solutions to the zero-sound frequency equation
and pion dispersion equation are considered as the functions
 of parameters  of the effective quasiparticle  interaction
 values (1).

1) We obtained the imaginary solutions
($\omega_P(k)$) to the zero-sound frequency equation (\ref{9})
responsible for the instability  of the ground state with respect to
 the long wave excitations. The branches $\omega_P(k)$ are obtained
for the different values of constants $f < 0$ of the effective
quasiparticle interactions.
 These branches lie on the unphysical
sheet  when $f$ satisfy the stability conditions (\ref{3}).
 However if   $f\leq -1/2,$ the imaginary branches
$\omega_P(k)$ go over to the physical sheet of the complex
frequency plane in some interval of momenta $(k_1,k_2)$. The value
of $k_2$ depends on $f$, $p_F$, $m^*$ while $k_1=0$.  This is a
signal about the instability of the ground state with respect to the
long-wave excitations.

2) The solution branches of $\omega^\pi_P(k)$ corresponding to
$\omega_P(k)$ are obtained for the pion dispersion equation
(\ref{6}).  It is shown that the
branches $\omega^\pi_P(k)$ pass to $\omega_P(k)$ if $f_{\pi
NN}\rightarrow 0$ . The solutions $\omega^\pi_P(k)$ give
instability at all negative values of $g'$.  When $g' \leq -1/2$
the instability emerges at  any density in correspondence of
(\ref{3}).  At $-1/2 < g' < 0$ nuclear matter becomes unstable at
some density larger the critical one:  $\rho > \rho_c$.
The value of density $\rho_c$ depends on $g'$ and other parameters
of the theory.

3)The four branches of solutions to the pion dispersion equation
(\ref{6}) with quantum numbers $J^\pi=0^-$ are obtained \cite{SR}.
Besides the well-known solutions: zero-sound, pion and isobar,  the
fourth branch $\omega_c(k)$ responsible for the pion condensation
is presented.

4)It is shown how the instability of the ground state
with respect to $\omega^\pi_P(k)$  is replaced by the instability
with respect to $\omega_c(k)$ when $g'$ changes near zero.
We cannot distinguish are the branches of solutions
$\omega^\pi_P(k)$ or $\omega_c(k)$ at $g' < 0$ on the physical
sheet.  However these branches are simply identified on the
unphysical sheet since they issue from different points:
$\omega^\pi_P(k=0)=0$ and $\omega_c(k=0)=m_\pi$. At small $k$
they are on the unphysical sheet. It is shown that at $g'=g'_{p}$
the part of $\omega^\pi_P(k)$ which is placed on the physical sheet
is replaced by the part of $\omega_c(k)$. This replacement is the
result of the interaction of three branches $\omega^\pi_P(k),$
$\omega_c(k)$ and $\omega^1_c(k)$ on the unphysical sheet.
Instability with respect to the solution belonging to the family
responsible for the long-wave instability is changed by instability
with respect to the "pion condensation".

5)The character of instability of the ground state with respect to
the long-wave excitations is considered. The zero-sound frequency
equation (\ref{9}) and the pion dispersion equation  (\ref{6}) are
symmetric with respect to the transformation  $\omega
\leftrightarrow -\omega$.  Therefore at $f \leq -1/2$ the
two imaginary solutions: $\omega_P(k)$ and -$\omega_P(k)$ appear
on the physical sheet. The both of them point out that the phase
transition took place. Applying the Migdal's selection rules
to $\omega^\pi_P(k)$ for the case of the large negative
$g'$, we observe that $\omega^\pi_P(k)$ which is lying on the
negative imaginary semiaxis belongs to ª $\pi^+$-type excitations.
This means that the corresponding  energies of physical excitations
have the negative imaginary parts. Then the ground state is
 unstable due to accumulation of
zero energy excitations in the ground state, but not because of the
exponential increasing of the any solution amplitudes.

\vspace{1cm}

The author is grateful to E.E.Saperstein for setting up a problem of
the stability conditions and M.G.Ryskin and E.G.Drukarev for
stimulating discussion.

This work was supported by the Russian Foundation for Basic
Research (project no.00-02-16853), DFG (project 438/RUS 113/595/0-1).

 \end{document}